# Does Cross-Field Influence Regional and Field-Specific Distributions of Highly Cited Researchers?

Xinyi Chen*

*Electrochemical Energy Reviews editorial office, Periodical Agency of Shanghai University, Shanghai 20444, China*
  *Coresponding author's Email address: shirleychen@i.shu.edu.cn

**Abstract**
Clarivate Analytics announces highly cited researchers (HCRs) every November to recognise true pioneers in their respective fields over the last decade who are one in 1,000 according to citation analysis based on the Web of Science™ database. However, the scientometric rules underlying HCR selections have constantly evolved over the years; thus, a comparative study between HCRs' academic relevance before and after 2018, when the cross field started to be included in HCR statistics, is essential. This paper evaluated the consistency of measurements in 2017 and 2018 by analysing HCR distributions by different regions and Essential Science Indicators (ESI) fields, studied the effects of introducing the cross-field category to the original 21 ESI fields, and portrayed the accurate picture of HCR distributions by region and subject without the influence of measurement biases. The cross field is believed to exert great impact on regional and field-specific HCR distributions, especially for research fields with HCR counts larger than 150. It was other countries and regions except the US and China that grew with the greatest momentum after the inclusion of cross-field HCRs.

## 1. Introduction
Citation databases are essential to supporting, discovering, accessing, and evaluating scholarly work. In recent years, many bibliometric indices, including the *h*-index (Hirsch, 2005) and its many variants, the highly cited publications (HCP) index (Bornmann, 2013), and altmetrics (Melero, 2015), that are based on citation databases have been introduced to evaluate the relationship between scientific impact and citations. Most of these indices focus on the scientific performance of individuals and do not classify researchers according to research field or set a threshold for each research field to identify researchers of "high impact." This has practical implications such as difficulties in evaluating and recognising individual researchers and limitations to assessing research excellence at the institutional, national, and regional levels (Waltman and Van Eck, 2012; Leskošek and Stare, 2015; Zhang, 2020).

It is interesting that compared with public institutions (such as national libraries), the private sector is more proactive in providing citation indices and offering bespoke services, with the former Science division of Thomson Reuters (which now belongs to Clarivate Analytics) being one of the most prestigious of such initiatives. The Highly Cited Researchers (HCRs) lists are constructed through the citation analysis of the well-acknowledged Web of Science™ database, which give clear indications of the world's scientists and social scientists over the last decade who are one in 1,000. The prototype of HCR rules was formulated in 2001, and two major variants of HCR rules have been put into practice since then, with the latest major variant that incorporated a cross-field category into the analysis in 2018.

This rolling HCR list, which is selected on a yearly basis since 2014, is a meaningful resource for bibliometric studies at the individual, institutional, and national levels. First, some countries in the global scientific community, such as China, are avoiding undue dependence on metrics such as journal impact factor (JIF) and the number of citations received in designing of remuneration schemes and employment of researchers, which leaves space for using HCR membership to illustrate individual academic performance instead of making direct reference to numbers of SCI-indexed papers or journal rankings in evaluations of researchers. Second, the HCR category took on significant additional meaning when the Essential Science Indicators (ESI) revealed the number of citations received by the top 1% institutions and researchers with a different baseline assigned for each different ESI field. Third, the lists also served as the basis for studies on personal and professional characteristics for talent management purposes (Amara *et al.*, 2015; Bornmann *et al.*, 2017; Parker, Lortie, and Allesina, 2010), and for bibliometric studies on social correlations between HCRs and institutional support and government funding (Docampo and Cram, 2019). Furthermore, the total number of HCRs at an institution not only is an indicator of research excellence of the institution with a weight of 20% in the well-known Academic Ranking of World Universities (ARWU) (Cai and Ying, 2005) but also provides valuable reference for administrative measures (Basu, 2006; Bornmann and Bauer, 2015).

As suggested by Garfield (Garfield, 1979), the citation count of papers, which is crucial in the bibliometric rule of HCR selections, is more of a measure of practical utility to researchers than an arguable indicator for academic research importance. Although this HCR list provides fertile and convenient data for research performance evaluation at different levels, there do exist a number of weaknesses. The first weakness is volatility of HCR membership. With revisions of the HCR rules, the methodology becomes more responsive to the emergence and ongoing success of preeminent researchers (Docampo and Cram, 2019). Emphasising emerging scientists has consequences for the volatility of HCR membership when considering its cross-year continuity, with approximately one-third to one-fourth of HCRs changing from year to year (Docampo and Cram, 2019). Such a variable HCR list presents an obstacle to connecting our general understanding of research excellence, as "the world's finest scholarly literature" (quoted from Clarivate[TM]) should be to a large extent contributed by time-honoured researchers and less so by those who briefly appeared in the HCR list. The second weakness is that there was a drastic increase in HCR volume within a short period from

2017 to 2018 when the methodology changed. With the third major variant of HCR rules in 2018, Clarivate Analytics nearly doubled the HCR sum worldwide within one year. However, few comparative studies on the consistency of HCR selection criteria and results before and after rule modifications in 2018 have followed, with only Docampo and Cram analysing the cross-year continuity of membership before 2017 (Docampo and Cram, 2019). The third weakness is that the comparability of HCRs across years is not ensured. For example, the cross-field HCRs in 2018 is suggested to achieve academic performance at the same level as HCRs from 21 other fields in 2018 according to the HCR selection algorithms (Clarivate, 2021). However, before setting the HCR standard of original 21 fields as the benchmark for cross-field HCRs of the same year, a prerequisite should be satisfied that HCRs from the 21 fields be certified as equally influential as their counterparts in the previous year; otherwise, the comparison between highly cited researchers across years and research fields would lose its validity. Thus, a comparative study of HCRs from 2017 and 2018 is especially crucial due to the literature gap and the latest major revisions to the selection rules introduced in 2018, which is not included in the methodology description on the website of Clarivate (Clarivate, 2021).

The robustness and steadiness of selection criteria for HCRs across years and research fields, especially when the model was undergoing transitions, are important factors that underpin talent management studies and administrative measures. This paper delves into the coherence of HCR ratings before and after the introduction of cross-field HCRs in 2018 to examine the methodological rigor of HCR selection rules and examines the influence of methodological differences on HCR profiles. More importantly, this paper paints an accurate picture of academic excellence worldwide and across disciplines which would have otherwise been neglected due to changes in methodology.

## 2. Methods

The lists of highly cited researchers from 2016 to 2020 examined in the present study are retrieved from https://recognition.webofscience.com/awards/highly-cited. The HCR data before 2016 are not included in this study because the second version of the HCR selection rules was in development from 2013 to 2015. Therefore, HCR products from 2016 to 2017 provided by Clarivate Analytics are mature and stable for representing the methodology of the second version of HCR rules, while the current HCR rules applied in 2018 represented the third version, which saw the addition of the cross-field category.

Then, a comparative study between the two sets of measurements was conducted in the present study. OriginPro 2018 (OriginLab Corporation, Northamptom, MA, USA) was used to analyse the categorical variables by Pearson's chi-squared test, conduct the nonparametric Wilcoxon signed-rank test, and test data distributions for normality. Excel computer spreadsheets (Microsoft, Seattle, WA, USA) were paramount for conducting a weighted least product (WLP) regression of the 2018 HCR profiles on their 2017 counterparts as well as constructing the 95% confidence intervals for the coefficients of the WLP regression and 95% limit of agreement (95% LOA) for regression results.

## 3. Evaluation processes
### *3.1 HCR populations from 2016 to 2020*

Fig. 1 shows the abrupt rise from 2017 to 2018 in the global HCR counts, as well as HCR counts for specific regions such as the US (with the highest share) and Chinese mainland (whose share has increased the most in recent years). Total HCRs in 2017 amounted to 3,538. That number increased to 4,059 in 2018, even excluding the cross-field category. In other words, the total number of HCRs from the original 21 ESI fields increased from 2017 to 2018. In addition to increased HCRs within these fields, the number of HCRs in the cross-field category reached 2,020 in 2018, which amounted to 6,079 HCRs in 2018 or 171% of the HCR total in 2017. Such a drastic increase in the global HCR total means cross-field HCRs had already existed for a long time but had not been captured by the HCR statistics. This conclusion puts into question the effectiveness and continuity of the selection rules for identifying HCRs before and after 2018.

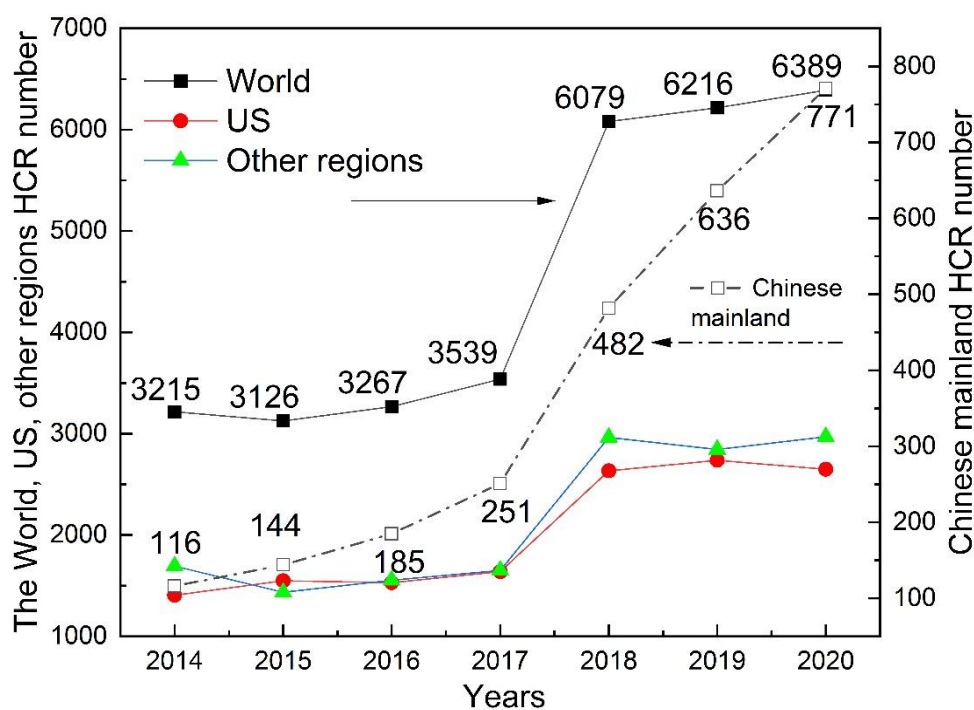

Figure 1 HCR counts for the world, US, Chinese mainland, and other regions from 2014 to 2020. Numerical labels of the data points show the HCR counts for the world and Chinese mainland.

In light of the conspicuous changes in the number of HCRs from 2017 to 2018 in contrast to its smooth trend from 2014 to 2017 and after 2018, the HCR counts of Chinese mainland over the years were investigated to briefly outline the relationship between the booming population of HCRs from Chinese mainland, which is the country showing the greatest momentum for HCR growth, and the years from 2016 to 2020. Pearson's chi-squared test of independence was used to assess if there was a significant relationship between two categorical variables ("being an HCR from Chinese mainland" and "different years from 2016 to 2020"), i.e., if proportion of HCRs from Chinese mainland dramatically changed over the years. Thus, we proposed two hypotheses:

$H_0$: The proportion of HCRs from Chinese mainland was not significantly related to the years.

$H_1$: The proportion of HCRs from Chinese mainland was significantly related to the years.

Three contingency tables (Tables 1 to 3) were made, of which the two variables are "being an HCR from Chinese mainland" and "from 2016 to 2017", "being an HCR from Chinese mainland" and "from 2017 to 2018", and "being an HCR from Chinese mainland" and "from 2018 to 2020," to separately assess the independence of "different years" on the other variable. Emphasis was placed on the time period "from 2017 to 2018," which encompasses the methodology changes in evaluating HCRs. We did not take into consideration data before 2016 because HCR rules changed from 2013 to 2015 and became fixed in 2016.

Table 1 Contingency table for Pearson's chi-squared test measuring whether the share of HCRs from Chinese mainland was independent of the time period from 2016 to 2017.

|  | 2016 | 2017 |
|---|---|---|
| An HCR from Chinese mainland | 184 | 251 |
| Not an HCR from Chinese mainland | 3,082 | 3,286 |

Pearson's chi-squared test data: chi-squared value = 6.07, d.f. = 1, *p value* = 0.014; $p < 0.05$; $H_0$ was rejected; there is significant evidence of an association between two variables.

Table 2 Contingency table for Pearson's chi-squared test measuring whether the share of HCRs from Chinese mainland was independent of the time period from 2017 to 2018 (during which the cross-field category was introduced).

|  | 2017 | 2018 |
|---|---|---|
| An HCR from Chinese mainland | 251 | 483 |
| Not an HCR from Chinese mainland | 3,287 | 5,596 |

Pearson's chi-squared test data: chi-squared value = 2.30, d.f. = 1, *p value* = 0.13; $p > 0.05$; $H_0$ was accepted; there is no significant evidence of an association between the two variables.

Table 3 Contingency table for Pearson's Chi-squared test measuring whether the share of HCRs from Chinese mainland was independent of the time period from 2018 to 2020.

|  | 2018 | 2019 | 2020 |
|---|---|---|---|
| An HCR from Chinese mainland | 482 | 636 | 769 |

| Not an HCR from Chinese mainland | 5,596 | 5,580 | 5,620 |

Pearson's chi-squared test data: chi-squared value = 57.56, d.f. = 2, $p < 0.001$; $H_0$ was rejected; there is significant evidence of an association between the two variables.

The sample size was larger than 40, and the expected cell counts in the contingency tables were greater than five, so the Pearson's chi-squared probability (the *p value*) was applied for interpretation. As shown in Tables 1 and 3, since the *p* values were less than 0.05, the null hypothesis that the two variables were independent from each other was rejected, and the accepted alternative hypothesis of Pearson's chi-squared tests in conjunction with Fig. 1 supports that there is a statistically significant relationship between the years and the share of HCRs from Chinese mainland from 2016 to 2017 and from 2018 to 2020, although not identified. However, as shown in Table 2, for the variables "being an HCR from Chinese mainland" and "from 2017 to 2018," the Pearson's chi-squared value $\chi^2 = 2.30$ ($p = 0.13$, $p > 0.05$) means the null hypothesis, which states that there was no significant evidence of association between the two variables, fails to be rejected. In other words, the proportion of HCRs from Chinese mainland changed drastically during 2017–2018 to the extent that the regular relationships, although not specified in the present study, between the years and the share of HCRs from Chinese mainland were broken up, thus changing the global geographic distribution of HCRs. Concurrently, the cross-field category was included in the citation analysis of HCRs; the concurrence is by no means fortuitous because no other significant factors intervened during 2017–2018. Thus, deeper exploration into populations of HCRs from different regions from 2017 to 2018 is needed to unveil the influence of the new methodology on the geographic distribution of HCRs.

*3.2 HCR populations for different regions*

Bridge charts for proportions of HCRs from different regions in 2017 and 2018 with and without cross-field HCRs clearly display the influence of introducing a new field category on demographics (Fig. 2). Compared with 2017, the share of HCRs from Chinese mainland increased to 251 or 7.93% in 2018 when including cross-field HCRs, while its share decreased to 6.80%, excluding cross-field HCRs, which is slightly down from accounting for 7.09% of the global share of HCRs in 2017. However, as the home for most HCRs in the world, the proportion of HCRs from the US would decrease in 2018 compared with 2017, down from 46.28% to 44.69%, without considering cross-field HCRs and further drop to 43.33% when including cross-field HCRs. Meanwhile, both percentiles increased in 2018 (48.51% when excluding cross-field HCRs and 48.74% when including cross-field HCRs) for "other regions except Chinese mainland and the US" (hereinafter referred to as "other regions"), up from 46.63% in 2017. In a word, Fig.2 illustrates the different impacts exerted by the cross-field category on different regions in a clear way.

Moreover, although the annual change of the proportion of HCRs from the US or other regions was not remarkable from 2017 to 2018, the trend was evident that the share of HCRs from the US decreased from 2014 to 2021 while that of other regions increased.

Specifically, the shares of HCRs from the US were 49.50% in 2015 and 39.70% in 2021, respectively. Trends like these are worth exploring to gain an accurate picture of global research excellence.

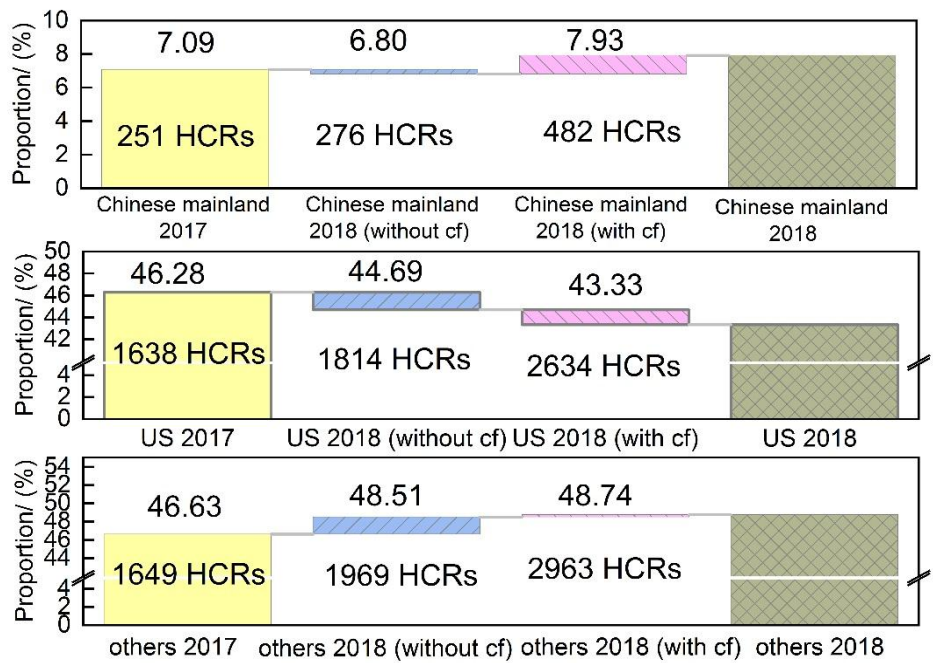

Figure 2 Bridge charts for the distribution of HCRs across regions in 2017 and 2018 including and excluding cross-field HCRs (denoted "cf" in the figure). Moreover, all percentage data corresponding to HCR counts are provided.

As illustrated in Fig. 3, the inclusion of the cross-field category in HCR ratings may also affect HCR distributions by the 21 ESI fields for different countries or regions. To analyse the distribution correlations between regions, four data pairs were included in the correlation analysis, which were the global, US, Chinese mainland, and other regions' HCR counts for 21 fields in 2017 and 2018. The normality test for each of eight datasets constituting the four data pairs using the Shapiro–Wilk method (suitable for sample sizes less than 50) revealed that at a significance level of 0.05, only other regions' HCR distribution in 2017 was significantly drawn from a normally distributed population among the eight datasets. Thus, the Spearman's correlation coefficient for ranked data is a relevant statistical tool for describing correlations between the eight HCR distributions that may not be normal. The resulting scatter matrix in Fig. 3 revealed positive correlations for every pair of data displaying significant correlations at a 0.05 level, except for one pair between the US and Chinese mainland, where a negative correlation existed in 2017 that became more significant in 2018. The distribution of HCRs from Chinese mainland displayed no remarkable correlations with the world or other regions, while the US and the other regions demonstrated significant positive relations with the world. These results reflect the fact that Chinese mainland had a set of HCR distribution behaviours across 21 fields that was distinct from that of other countries or regions. Notably, enhanced intensity of negative correlation between Chinese mainland and the US, and the strengthened significant positive correlations between the US and the world and between other regions and the world consistently indicated that HCR distributions among different regions had become more multi-polarized in 2018 compared to 2017.

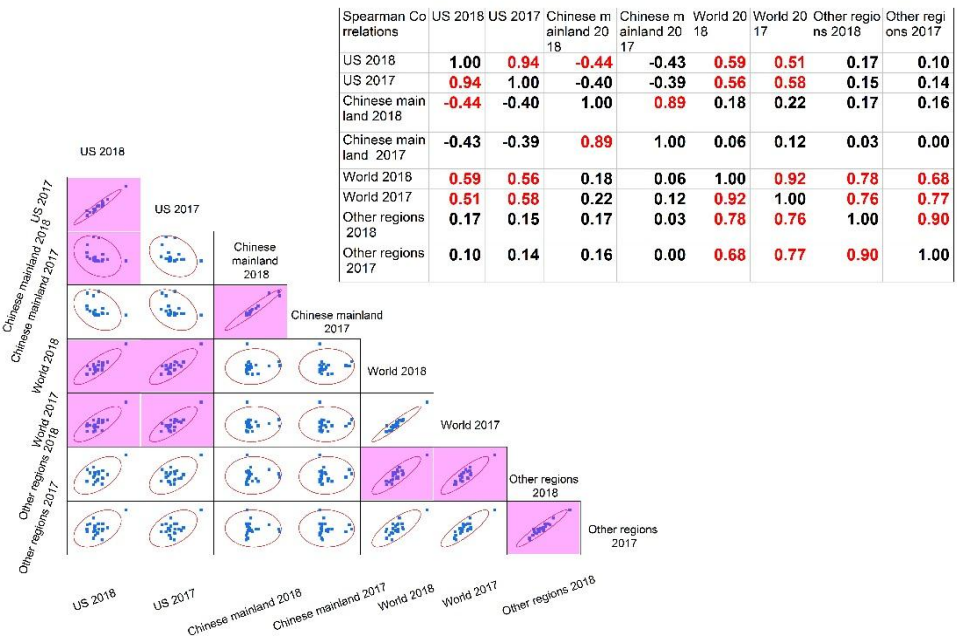

Figure 3 Scatter matrix showing the Spearman correlations among HCR distributions by the 21 fields for the world, US, Chinese mainland, and other regions in 2017 and 2018, for a total of eight HCR distributions. Confidence ellipses at the 0.05 significance level are added. Cells with a background colour of magenta demonstrate significant correlations at a level of 95%. The top-right table shows the Spearman correlation coefficients corresponding to the scatter matrix, with significant correlation values marked in red for reference.

*3.3 HCR populations for 21 ESI fields*

According to the 2018 HCR rule (Clarivate, 2021), Clarivate does not specify what impact that dividing the whole scientific population into 22 fields would have on the original HCR distribution by the 21 fields. Thus, comparing HCR distributions before and after considering the cross-field HCRs serves as the basis for evaluating the effect of the new methodology applied in 2018 and its interchangeability with the old methodology. To that end, eight groups of data from the scatter matrix, which are HCR counts for 21 fields in 2017 and 2018 for the whole world, US, Chinese mainland, and other regions, would be treated as four pairs of samples in a difference test. A paired difference test was sought for comparing two sets of measurements to assess whether their population means differed. In light of the above mentioned nonnormality of HCR distributions by 21 fields in 2017 and 2018 for different regions, the Wilcoxon signed-rank test specifically designed for differences that may not be normally distributed was carried out to analyse the influence imposed by the new methodology upon 21 subjects for different regions.

The Wilcoxon signed-rank test results showed that at a significance level of 0.05, only the two distributions of HCRs by 21 fields for Chinese mainland in 2017 and 2018 were not significantly different, in contrast to the significant differences within the other three pairs of distributions, which were global HCRs, US HCRs, and other regions' HCRs, by 21 fields in 2017 and 2018. These results indicate that for Chinese mainland, the updated measurement in 2018 was statistically equivalent to the former in 2017 at the 0.05 level with regard to HCR distributions by 21 fields, while the two measurements led to nonequivalent effects on those distributions for the world, US, and other regions.

What we learned from the above results is that the two methodologies for HCR selections in 2017 and 2018 could have affected the HCR distributions by regions or ESI fields in different manners. However, merely determining whether there were differences or inconsistencies between the two sets of measurements would not provide adequate theoretical foundations for answering a variety of methodology comparison questions related to the calibration of bias or the range of differences between the two methods. These considerations will be elaborated in the next section.

*3.4 Methodological assessment*

The regression of HCR counts obtained from method two (M2, i.e., the methodology in 2018) on those from method one (M1, i.e., the methodology in 2017) can provide informative results for comparing the two methods and is more straightforward in demonstrating the measurement biases. Moreover, the conformity to normal distributions of HCRs by regions and research fields in 2017 and 2018, which was tested with the Kolmogorov–Smirnov method (which is suitable for the normality test of a sample size larger than 50), also supported the construction of regression on the HCR counts.

Systematic biases, such as fixed biases and proportional biases that would carry distinct physical significances, can be separately addressed by constructing the weighted least product (WLP) line of best fit around the original scattered data using the method explained by Ludbrook (Ludbrook, 2002; Ludbrook, 1997; Ludbrook, 2010), who suggested that this method could cater to cases where random errors are attached to both dependent and independent variables. Therefore, we conducted WLP regression analysis of the original data (see Fig. 4), with HCR counts of 21 fields in 2018 as the $y$-axis (dependent) and those in 2017 as the $x$-axis (independent), to express the size of the measurement biases in quantitative terms.

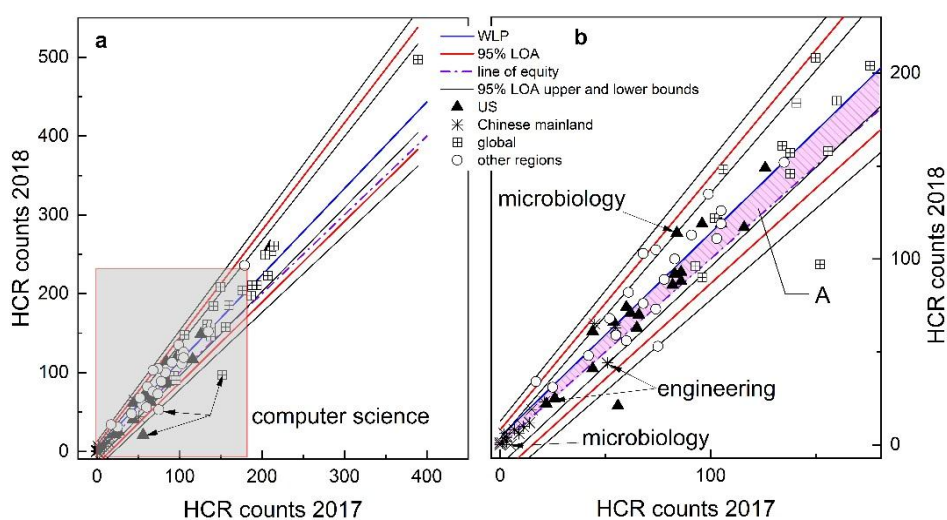

Figure 4 Weighted least product regression analysis of **a** HCR counts in 2018 vs. those in 2017; and **b** enlarged view of the grey area in (a) showing the part of dense observations, with the area in violet colour being where HCR counts that saw growth in the ESI fields in 2018 compared with 2017 actually decline relative to the WLP regression line. Annual growth of the scientific community was overlapped with systematic bias of the 2018 measurement, and the annual growth corresponded to the violet area in (b) and would be termed "A."

The coefficients of the WLP regression line of best fit for HCR counts in 2018 (method 2, i.e., M2) vs. HCR counts in 2017 (method 1, i.e., M1) are calculated by the Excel spreadsheet, and the result is shown in Eqn. 1:

$$1.10\ M1 + 4.20 = M2 \qquad \text{Eqn. 1}$$

of which the critical assumption that the residuals around the regressions are normally distributed is satisfied at the 0.05 significance level, showing the effectiveness and goodness-of-fit for the fitted WLP model to the observed values.

The 95% CIs for the slope and the intercept were 0.29~7.94 and 1.05~1.14, respectively; for comparison, the 45º line of equity was drawn to represent the case of identical outcomes from the two methods. Thus, the following statistical inferences can be made: (1) The hypothesis that the intercept equals zero (i.e., no fixed bias between the methods) was rejected at the 0.05 significance level because the 95% CIs for the intercept did not include zero; and (2) The hypothesis that the slope equals one (i.e., no proportional bias between the methods) was rejected at the 0.05 significance level because the 95% CIs for the intercept did not include one. In other words, M2 gave HCR counts distributed by 21 fields and different regions that not only were progressively higher than M1 by approximately 10% but also definitely contained a fixed increment (approximately 4.20).

The quantitative analysis of both proportional and fixed biases cannot provide a benchmark against which to determine whether the magnitude of measurement biases allowed M2 to substitute for M1 in analysing HCR distributions by 21 fields because the systematic bias detected could stem from natural variability in the HCR population from 2017 to 2018. To draw the disagreement of two methods from the systematic biases, the method agreement assessment would incorporate qualitative factors within quantitative analysis, i.e., referring to the limit of agreement (LOA) as a supplement to the regression analysis (Bland and Altman, 1999; Woodman, 2010) to determine if the two methodologies are interchangeable in analysing HCR distributions by 21 ESI fields. Thus, 95% confidence limits were constructed by adapting the Bland technique (Bland, 2006) to allow building of V-shaped confidence limits around the WLP regression line. The steps are as follows:

1. Extract the residuals as the differences between the observed and fitted values for Eqn. 1;
2. Convert the residuals into absolute values;
3. Construct the ordinary least squares (OLS) regression of absolute residuals on M1, where

$$\text{Absolute residual} = a + b \times M1 \qquad \text{Eqn. 2}$$

The coefficients are given in Table 4.

Table 4 Outcomes of OLS regression of absolute residuals on M1. The confidence level for parameters is 95%.

|  | Value | Standard Error | *t* Value | Prob. > \|*t*\| |
|---|---|---|---|---|
| Intercept (*a*) | 4.85 | 1.88 | 2.58 | 0.012 |
| Slope (*b*) | 0.07 | 0.02 | 3.99 | < 0.001 |

4. Adjust the coefficients of the regression in Eqn. 2 by multiplying them by $\sqrt{(\pi/2)} = 1.25$. This is because the mean of the absolute value of a normal distribution with mean zero and standard deviation $\sigma$ is $\sqrt{(2/\pi)} \times \sigma$. Then, we obtain:

$$SD = 1.25a + (1.25b) \times M1 \qquad \text{Eqn. 3}$$

5. Thus, the 95% limits of agreement (95% LOA) of the M2 value for a given M1 value are:

$$\text{Fitted value} \pm 1.96\, SD \qquad \text{Eqn. 4}$$

where the fitted value is obtained from the regression Eqn. 1, and 1.96 is the standardised normal deviate corresponding to the two-sided $p = 0.05$.

It should be noted that Ludbrook (Ludbrook, 2010) preferred using $t_{0.05, d.f. n-1}$ to using 1.96 to account for sample size adjustment. In the present study, the sample size is 84, which corresponds to $t_{0.05, 83} = 1.99$. Thus, both techniques give almost equal results.

Now, we obtain the 95% confidence limits for the upper and lower "limits of agreement" (95% LOA) for the WLP regression results, as shown in Fig. 4 (a), and the results were theoretically an indication of interchangeability of M1 and M2 in terms of analysing HCR distributions by 21 fields and different regions in that 95% LOA included 95% of observations. At this point, the variation within a set of measurements for the fitted values was attended by SD, but SD was estimated based on absolute residuals, which

were subject to measurement errors represented by $SE_a$ and $SE_b$ in Table 4. Thus:

$$SD_{total} = 1.25\,(a \pm SE_a) + 1.25\,(b \pm SE_b) \times M1$$
$$= 1.25\,(a + b \times M1) + 1.25\,(\pm SE_a \pm SE_b \times M1)$$
$$= SD + 1.25\,(\pm SE_a \pm SE_b \times M1) \qquad \text{Eqn. 5}$$

That is, a measurement error term $1.96 \times 1.25 \times (\pm SE_a \pm SE_b \times M1)$ for the M2 value with a given M1 value, which would be either added to the upper limit of 95% LOA or subtracted from the lower limit of 95% LOA to construct the estimated upper and lower bounds around 95% LOA. The error term representing measurement repeatability (the residual variations) implies how proportional or fixed errors from repeated measurements can affect the estimates of 95% LOA of the parent populations. The measurement errors increased in proportion to the level of M1, as shown in Fig. 4, and when the HCR counts were larger than 150, the estimated 95% LOA would be seriously affected by the error term, as demonstrated by the intersection of the line of equity and the upper bound of the lower limit of 95% LOA, implying that the new methodology applied in 2018 for analysing HCR distributions by 21 fields would safely substitute for the old methodology applied in 2017 if the HCR counts for the studied ESI fields were smaller than 150 but may more seriously affect the observations with HCR counts larger than 150 to the extent that M1=M2 would not be included in the 95% LOA. In other words, methodological biases for ESI fields with HCRs greater than 150 seemed to be the rule rather than the exception by transforming the HCR selection rules to the 2018 version.

## 4. Discussion

*4.1 Measurement inconsistency between methodologies in 2018 and 2017*

In 2018, Clarivate updated its methodology to select HCRs from 21 fields plus a cross-field, and the inclusion of cross-field HCRs exerted significant impacts on HCR distributions by region and research field. WLP regression analysis showed that there were both proportional and fixed biases between the two sets of measurements. Thus, it is important to differentiate the influence of the updated methodology from natural variations in the HCR group characteristics from 2017 to 2018. The analysis of 95% LOA of the WLP regression revealed that the two methodologies can be interchanged in HCR analysis for 21 fields when the HCR counts of the sample fields were less than 150, while the methodological bias was so intense for observations greater than 150 on account of measurement errors that the new methodology cannot "safely" substitute for the old one as the HCR counts surpassed 150. The Wilcoxon signed-rank test also corroborated the above inferences in that HCR distributions for Chinese mainland in 2017 and 2018 were statistically the same because HCR counts for Chinese mainland in 21 fields were within the 95% LOA and no field was above 150, synergistically ensuring the agreement of two methodologies for Chinese mainland. The US, other regions, and the whole world all had several outliers with respect to 95% LOA and multiple observations greater than 150, which resulted in differences between HCR distributions for these areas in 2017 and 2018 as supported by statistical significance in the Wilcoxon signed-rank test.

*4.2 Influence of the cross-field on 21 fields*

If cross-field HCRs were excluded from 2018 data, then both the US and Chinese mainland would see drops in their proportions of HCRs (Fig. 2). The results correspond to many observations below the WLP regression line (Fig. 4 (a)), which suggests that observations that are negatively affected by the introduction of the cross-field HCRs, are responsible for the decreased proportions of HCRs from the US and Chinese mainland in 21 fields in 2018.

We can obtain fields that generate cross-field HCRs from the further analysis of negative influence of the cross-field on these observations. The line of equity in Fig. 4(a), which clearly divided observations into two groups and intuitively served as the dividing line between measurements with values $M_1 > M_2$ and $M_2 > M_1$, means that $M_1 = M_2$. Only a few observations lay in the area where $M_1 > M_2$, indicating that most research fields underwent conspicuous growth from 2017 to 2018. However, annual growth of the scientific community was overlapped with systematic bias of the 2018 measurement, and the annual growth corresponded to the violet area in Fig. 4 (b) and would be termed "A," while the systematic bias was shown by the WLP regression line. The WLP regression line signifies the influence of the cross-field category (denoted "cf.") on the 21 fields. The area above the WLP line shows positive influence of the cross-field category on the HCR counts, i.e., cf. > 0, while the area below the WLP line shows negative influence of this category, i.e., cf. < 0.

Regarding the observations below the WLP line, the distance between these observations and the line of equity shows measurement bias between 2018 and 2017 for these fields, i.e., the difference between the annual growth and the absolute value of cf., or A−|cf.| (cf. is negative). The cf. contains the influence of all years of cross-field HCRs before 2018, and thus may be larger than A, while the annual cross-field HCRs generated from each field should be smaller than A. As shown in Fig. 4(a), computer science (A<|cf.|) has been continuously forming interdisciplinary fields with other fields before 2018.

The area below the WLP line but above the line of equity, shown in violet in Fig. 4 (b), shows A > |cf.|. The physical meaning of the violet area is clear: although these observations displayed an apparent annual increase, they decreased with respect to the WLP line. The annual growth of the observations in the violet area was negatively influenced by the cross-field category in 2018 thus generating cross-field HCRs. Because Clarivate identified HCRs of the cross-field category in 2018 based on their citation scores of all fields in 2018, the generated cross-field HCRs from these fields were associated with the citations from cross-fields in 2018, thereby making these fields cutting-edge knowledge-generating fields. The area includes observations such as environment/ecology, immunology, psychiatry/psychology, social science and chemistry (only for the US); computer science and physics (only for Chinese mainland); space science, immunology, neuroscience and behaviour, and agricultural sciences (only for other regions); economics and business, immunology, physics, plant and animal science, neuroscience and behaviour, and social science (only for the whole world), which were believed to be fields that generated cutting-edge sciences and technologies for these countries or regions in 2018.

Regarding the area above the WLP line, these observations faced a short-term increase even when including the cross-field category, as opposed to decreasing HCR counts with regard to those observations below the WLP line. Thus, the positive influence of the cross-field HCRs on the 21 fields means these observations generated fewer cross-field HCRs during 2017–2018 than those observations below the WLP line.

This revelation would provide valuable references for talent management and the journal publishing industry.

*4.3 Accurate picture of research excellence worldwide and its implications*

(1) *Original 21 ESI fields.* The detected negative correlations between the US and Chinese mainland in Fig. 3 are also visually verified in Fig. 4, in which the US saw the greatest decline in computer science among all research fields in 2018, but Chinese mainland did not experience such a drastic decrease. Chinese mainland witnessed the greatest drop in microbiology and engineering among all fields, but the US performed better in these two fields, especially in microbiology showing a substantial increase with respect to the WLP line, which outperformed almost all other fields in the US case. The strengthened negative correlations between Chinese mainland and the US in 2018 should have other origins than the updated HCR selection rules in 2018 because the 2018 method did not favour Chinese mainland at the expense of the US in terms of the share of HCRs in the 21 fields. The share of HCRs from Chinese mainland in the 21 fields in 2018 decreased by more than that from the US, from 7.09% to 6.80% (dropping by 4.1%) vs. from 46.28% to 44.69% (dropping by 3.4%).

Conversely, other regions saw a rise in their global share of HCRs in 21 fields from 2017 to 2018, which was a strong indication of powering ahead and may account for and directly lead to a relative fall of the share of HCRs from the US in these fields.

(2) *The cross field.* As science and technology are developing, an increasing number of research fields are forming. The cross-field category constitutes nearly one-third of all HCRs; moreover, highly cited papers exhibit interdisciplinary patterns (Chen *et al.*, 2021). Therefore, a more detailed subdivision between different cross-fields can reflect the advances in science and technology. According to the present study, computer science had been continuously forming interdisciplinary fields with other fields before 2018; e.g., artificial intelligence (AI) is the cross-field of computer science, psychology and philosophy, and is not a new cross-field. But AI remains a sub-category of computer science in the JCR 2021, which would be a hurdle for the development of AI since its multi-disciplinary feature is not receiving enough attention.

Furthermore, establishing more specific research fields in ESI/JCR for measuring the academic impact of scientists from the cross-field also better serves the ever-growing scholarly research of more innovative talent. Still take AI as an example, the US Department of Energy announced several new developments in the department's ongoing efforts to advance AI in 2019. Peking University also established the Institute for Artificial Intelligence in 2019.

(3) *Implications of the accurate picture.* Chinese mainland pursued its own path for HCR growth because it did not show any significant positive correlations with other parts of the world. HCR distributions among different regions became more multi-polarized in 2018 than in 2017 according to the correlation analysis. However, the multi-polarized feature may be overlooked by Clarivate Analytics in its Highly Cited

Researchers 2020 report (Clarivate, 2020), which publicised the notion that Chinese mainland roars ahead while the US lags in the shares of HCRs as research has globalised. This notion should be questioned since it was other regions who increased the share of HCRs in the context of both Chinese mainland and the US decreasing theirs from 2017 to 2018 (Fig. 2). These regions include many developed Western countries who rank high in the HCR lists by country or region. They were potential competitors for the US, because the positive correlations among the world, US, and other regions regarding their HCR distribution patterns in the 21 fields indicated that these countries/regions would encounter fewer difficulties in following the US' mode to improve their global shares of HCRs. However, competitions arising from cultural similarities fail to draw adequate attention compared with those that could be associated with "cultural conflict" between Western and Eastern countries (Huntington, 1996).

Cultural conflict is irrelevant to the evolution of global research excellence. However, failing to understand this point may result in negative repercussions in global cultural exchanges, e.g., the controversial China Initiative launched in 2018 by the US government.

*4.4 Limitations of the present study and future perspectives*

The present study has limitations in that the population of HCRs increased by approximately 14.73% over one year from 2017 to 2018 in terms of the 21 ESI fields, which precluded translating the size of measurement biases into the magnitude of disagreement between the two methods. However, it is not the intention of the present study to judge whether the two methods are interchangeable according to the size of method differences, although the inconsistency of the HCR selection methodologies is discussed. The updated selection rules with the cross-field category have already become an inalienable part of the HCR lists. Thus, it is important to detect and analyse biases to paint an accurate picture of research excellence with respect to research fields and regions excluding the influence of measurement biases, which was clearly put forward in Sections 4.1 to 4.3.

The present study can serve as the foundation for future comparative research on talent quality between regions and research fields since 2018 was the first year in which cross-field HCRs were considered. Thus, the statistical regularity demonstrated by using the data from 2017 to 2018 can better account for inconsistent variations in countries and research fields than by using data from other years, e.g., 2017 data representing the old methodology and 2021 data representing the new methodology, because using data from 2017 to 2018 ensures that the observed inconsistent variations may to the least extent be caused by longer time of HCR evolution but be largely due to the method transformation. Moreover, deviant behaviours would follow the universal recognition of cross-field HCRs which brought about a sharp rise in the global HCR total, including institutions seeking quick ranking promotion by recruiting more HCRs and growth of guest authorship, which would distract education policies from responding to reality, exacerbate inequality of resource allocations among regions and disciplines, eventually skew the statistics of HCRs. Thus, the HCR lists from 2017 to 2018 are the most meaningful resources for bibliometric studies to capture the academic behaviour of HCRs.

In summary, Clarivate directly uses the changing number of HCRs to represent the changing research excellence among different regions and research fields, but doesn't mention the effect of HCR selection rules. Fortunately, the introduction of the cross-field HCRs is the silver lining; we can differentiate the changes effected by the cross-field from the natural variations in the HCR group characteristics during 2017–2018. This paper is valuable in that it illustrates the shift in scientific excellence as a result of methodological modifications in 2018 and reveals a more reliable and clearer picture of research excellence worldwide, which would otherwise be overlooked and consequentially influenced by artificial factors such as economic potential in terms of GDP per capita (Benito *et al.* 2020). Future investigations would focus on HCR distributions by country, especially for other parts of the world expect the US and Chinese mainland to better illustrate the global landscape of HCRs on a country-by-country basis.

## 5. Conclusion

HCR lists are informative and valuable resources for scientific research management and for society at large. Thus, consistent HCR selection rules are essential to the sustainable development of scientific and educational undertakings. HCR distributions in 2017 and 2018 were studied to investigate the consequences of changing methodologies on different regions and research fields, and statistically significant biases above HCR counts of 150 were detected for the 2018 method. Additionally, although the 2018 HCR lists saw substantial growth in HCR counts for most ESI fields and regions, some of these observations encountered actual drops relative to the WLP line, which indicates research fields that were generating cutting-edge knowledge. Other regions except the US and Chinese mainland witnessed increased shares of HCRs regardless of whether the cross-field HCRs were included, implying great momentum of growth. The inconsistent variation in the regional and field-specific distributions of HCRs from 2017 to 2018 described in this study would have otherwise been overlooked. This study could serve as a starting point for future research regarding the impact of cross-field HCRs on different regions and subjects and provide valuable references for talent management and the journal publishing industry that are based on the ESI HCR lists.

Informetrics, 14(3), 101038.